\documentclass{article}        

\usepackage[latin1]{inputenc}
\usepackage{graphicx}
\usepackage{amssymb}
\usepackage{amsmath}
\usepackage[super]{natbib}
\usepackage{marvosym}
\usepackage{multicol}
\usepackage{epsfig}
\usepackage{psfrag}
\usepackage[labelfont=bf]{subfig}
\usepackage[english,ruled,vlined]{algorithm2e}

\bibpunct{(}{)}{,}{a}{}{;}

\newcommand{\cst}{\mathrm{cst}}

\newcommand{\E}{\mbox{$\mathsf E$}}

\numberwithin{equation}{section}

\DeclareMathOperator{\Z}{{\bf Z}}
\DeclareMathOperator{\n}{{\bf n}}
\DeclareMathOperator{\X}{{\bf X}}
\DeclareMathOperator{\KL}{KL}
\DeclareMathOperator{\Dirichlet}{Dir}
\DeclareMathOperator{\Beta}{Beta}
\DeclareMathOperator{\balpha}{\boldsymbol{\alpha}}
\DeclareMathOperator{\bpi}{\boldsymbol{\pi}}
\DeclareMathOperator{\boeta}{\boldsymbol{\eta}}
\DeclareMathOperator{\bozeta}{\boldsymbol{\zeta}}
\DeclareMathOperator{\btau}{\boldsymbol{\tau}}

\begin{document}

\title{Variational Bayesian inference and complexity control for stochastic block models}

\author{P. Latouche\thanks{Address for correspondance: Pierre Latouche, Laboratoire Statistique et G\'enome, Tour Evry 2, 523 place des terrasses de l'Agora, 91000 Evry, France.
E-mail: pierre.latouche@genopole.cnrs.fr} , E. Birmel\'{e} and C. Ambroise \\
Laboratoire Statistique et G\'enome, UMR CNRS 8071, UEVE
}
\date{}

\maketitle


{\bf Abstract:}
  It is now widely accepted that knowledge
  can be acquired from networks by clustering their vertices according to
  connection profiles.  Many methods have been proposed and   in this
  paper  we concentrate  on   the Stochastic  Block Model (SBM).  
The clustering of vertices and the estimation of SBM
  model parameters have been subject to previous work and  numerous
  inference strategies such as variational Expectation Maximization (EM) and classification EM
  have been proposed. However, SBM still suffers from a lack of criteria to
  estimate the number of components in the mixture.
To our knowledge, only one model based criterion,
  ICL,  has been derived for SBM in the literature. It relies on
  an asymptotic approximation of the Integrated Complete-data
  Likelihood and recent studies have shown that
  it tends to be too conservative in the case of small networks.
  To tackle this issue, we propose a new criterion that we call
  ILvb, based on a
  non asymptotic approximation of the marginal likelihood. We describe how the criterion can be computed through a
  variational Bayes EM algorithm. 
\medskip

\medskip

Received: date / Accepted: date


\section{Introduction}
\label{Section:introduction}

Networks  are   used  in  many  scientific  fields   such  as  biology
\citep{Article:Albert:2002} and social
sciences \citep{Article:Snijders:1997,Article:Nowicki:2001}. They aim at modelling with
edges  the way  objects  of interest  are
related to each other. Examples of such data sets are friendship
\citep{Article:Palla:2007},
protein-protein  interaction  networks \citep{Article:Barabasi:2004},
powergrids     \citep{Article:Watts:1998}   and     the    Internet
\citep{Article:Zanghi:2008}. In this context, a
lot of attention has been paid on developing models to learn knowledge
from the network  topology. It appears that  available  methods  can
be grouped into three significant categories.

Some  models look for  community structure,  also called  homophily or
assortative mixing \citep{Article:Girvan:2002,Article:Danon:2005}. Given a network, the vertices are partitioned into
classes such that vertices of a class are mostly connected to vertices
of  the  same  class.   In the model of
\cite{Article:handcock:2007}, which extends \cite{Article:hoff:2002}, vertices  are clustered depending
on their positions in a continuous latent space. They proposed a
two-stage maximum likelihood approach and a Bayesian algorithm, as
well as an asymptotic BIC criterion to estimate the number of
latent classes. The two-stage maximum
likelihood approach  first maps the  vertices in the latent  space and
then uses a  mixture model to cluster the  resulting positions.  In
practice, this
procedure converges quickly but looses  some information by not estimating  the positions and
the  cluster  model  at  the  same  time.   Conversely,  the  Bayesian
algorithm, based on Markov Chain Monte Carlo, estimates both the latent positions and the mixture model parameters
simultaneously. It  gives better results  but is time  consuming. Both
the maximum  likelihood and the  Bayesian approach are  implemented in
the R package ``latentnet'' \citep{Manual:Krivitsky:2009}.

Other models look for disassortative mixing, in which vertices mostly
connect       to       vertices       of       different       classes
\citep{Article:Estrada:2005}.       They      are
particularly suitable for the analysis of bipartite networks which are
used  in numerous  applications.  Examples  of data  sets  having such
structures are
 transcriptional regulatory networks where operons encode transcription factors directly involved in operons regulation. To get some insight into the transcription
process, these two types of nodes are often grouped into different
classes with high inter connection probabilities. Other examples are citation networks where authors cite or are cited by papers.
For a  more detailed description of the  differences between community
structure and disassortative mixing, see \cite{Article:Newman:2007}.

Finally, a few models can look for both community structure and disassortative mixing. \cite{Article:hofman:2008}  proposed  a 
probabilistic   framework,  as   well  as   an   efficient  clustering
algorithm.  Their model, implemented in the software ``VBMOD'', is based on two key parameters $\lambda$ and
$\epsilon$.  Given a network, it assumes that vertices connect with probability
$\lambda$   if  they  belong   to  the   same  class   and  with
probability $\epsilon$
otherwise. Moreover, they introduced a non asymptotic Bayesian criterion
to estimate the number of classes. It is based on a variational approximation
of the marginal likelihood  and  has  shown   promising  results. 
In this paper, we focus on the Stochastic Block Model (SBM) which was  originally developed in
social sciences
\citep{Article:White:1976,Article:Fienberg:1981,Article:Frank:1982,Article:Holland:1983,Article:Snijders:1997}.
Given  a  network,  SBM assumes  that  each
vertex
belongs to a hidden class among $Q$ classes, and uses a $Q \times Q$ matrix $\bpi$ to
describe the intra and inter connection probabilities. Moreover, the
class proportions are represented using a $Q$-dimensional vector $\balpha$. No  assumption is  made on the  form of
 the connectivity matrix such that  very
different structures can be taken into account.  In particular, SBM can characterize
the presence
of hubs which make networks locally dense \citep{Article:Daudin:2008}. Moreover and to some extent, it generalizes many of the existing graph clustering techniques as  shown in  \cite{Article:Newman:2007}. For  instance, the  model of
\cite{Article:hofman:2008} can be seen  as a constrained SBM where the
diagonal of $\bpi$  is set to $\lambda$ and all  the other elements to
$\epsilon$.

Many methods have been proposed  in the literature to jointly estimate SBM
model parameters and cluster the vertices of a  network. They all face the
same difficulty.  Indeed,
contrary to many mixture models, the conditional 
 distribution of all the latent variables $\Z$ and model parameters, given the observed data $\X$, can not
be        factorized due to conditional dependency       (for        more        details,        see
\citealt{Article:Daudin:2008}). Therefore, optimization techniques such as the
EM algorithm can not be used directly.  \cite{Article:Nowicki:2001} proposed a Bayesian probabilistic
approach. They  introduced some prior Dirichlet  distributions for the
model  parameters and used  Gibbs sampling  to approximate  the posterior
distribution over the model parameters and posterior predictive distribution. Their algorithm is implemented in the software BLOCKS,
which is part of the package StoCNET  \citep{Manual:Boer:2006}.  It gives accurate a posteriori
estimates but can  not handle  networks with
more than 200 vertices. \cite{Article:Daudin:2008}
proposed  a frequentist  variational  EM approach for SBM
which can
handle much larger networks.  Online   strategies    have   also   been   developed
\citep{Article:Zanghi:2008}.  

While many inference strategies have been proposed for estimation and
clustering purpose, SBM still suffers from a lack of criteria to
estimate the number of classes in networks. Indeed, many criteria, such as the  Bayesian Information
Criterion   (BIC)   or   the   Akaike  Information   Criterion   (AIC)
\citep{Book:Burnham:2004} are based  on the likelihood $p(\X|
\balpha, \bpi)$ of the observed data
$\X$,     which    is     intractable     here.     To tackle this issue,
\cite{Article:Mariadassou:2010} and  \cite{Article:Daudin:2008}
used a  criterion, so-called ICL,  based on  an
asymptotic approximation of the integrated
\emph{complete-data} likelihood. This criterion relies on the joint distribution
$p(\X, \Z|\balpha, \bpi)$ rather than $p(\X|\balpha, \bpi)$ and can be
easily computed, even in the case of SBM. ICL was
originally proposed by \cite{Article:Biernacki:2000} for model
selection in Gaussian mixture models, and is known to be particularly
suitable for cluster analysis view since it favors well separated
clusters. However, because it relies on an asymptotic approximation,
 \cite{article:Biernacki:2010} showed, in the case of mixtures of
 multivariate multinomial distributions, that it may fail to detect interesting structures present in the
 data, for small sample sizes. \cite{Article:Mariadassou:2010}
 obtained similar results when analyzing networks generated using SBM. They found that this asymptotic criterion  tends
 to underestimate the number of classes when dealing with small networks.
 We emphasize that, to our knowledge, ICL is currently the only \emph{model based
  criterion} developed for SBM.

%
%
%

Our main concern in this paper is to propose a new criterion for
SBM, based on  the marginal
likelihood $p(\X)$, also called integrated
\emph{observed-data} likelihood. The marginal likelihood is known to focus on density
estimation view and is expected to provide a consistent estimation of
the distribution of the data. For a more detailed overview of the
differences between integrated \emph{complete-data} likelihood and
integrated \emph{observed-data} likelihood, we refer to
\cite{article:Biernacki:2010}. In the case of SBM, the marginal
likelihood is not tractable and we describe in this paper
how a non asymptotic approximation can be obtained through a
variational Bayes EM algorithm.

In Section \ref{Section:mixtureModel}, we describe SBM and we introduce some non
informative conjugate prior distributions for the model
parameters. The variational Bayes EM algorithm is then presented in Section
\ref{Section:estimation}. We show in Section \ref{Section:modelSelection} how it naturally leads to a new
model selection criterion that we call ILvb, based on a non asymptotic approximation of
the marginal likelihood.
Finally, in Section \ref{Section:experiments}, we carry out some experiments using simulated data sets and the metabolic network of
\emph{Escherichia coli}, to assess ILvb.

\hspace{-1.5em}The R package ``mixer'' implementing this work is available
from the following web site:
\verb#http://cran.r-project.org#.

\section{A Mixture Model for Graphs}
\label{Section:mixtureModel}
The data we model consists  of a $N \times N$ binary matrix  $\X$,
with entries $X_{ij}$ describing the presence or absence of an edge
from vertex $i$ to  vertex $j$. Both directed and undirected relations
can be analyzed but in the following, we focus on undirected
relations.  Therefore $\X$ is symmetric.

\subsection{Model and Notations}
The Stochastic Block Model (SBM) introduced by \cite{Article:Nowicki:2001} associates to each vertex of a
network a latent variable $\Z_{i}$ drawn from a multinomial distribution, such
that $Z_{iq}=1$ if vertex $i$ belongs to class $q$
\begin{displaymath}
  \Z_{i} \sim \mathcal{M}\Big(1, \:\balpha =
  (\alpha_{1}, \alpha_{2}, \dots, \alpha_{Q})\Big).
\end{displaymath}
We   denote  $\balpha$,   the   vector  of   class
proportions. The edges are then drawn from Bernoulli distribution 
\begin{displaymath} 
  X_{ij} |\{Z_{iq}Z_{jl} = 1\} \sim \mathcal{B}(\pi_{ql}),
\end{displaymath}
where $\bpi$ is a $Q \times Q$ matrix of connection probabilities.
According to this  model, the latent variables $\Z_{1}, \dots,
\Z_{N}$ are iid and given this latent
structure, all  the edges are  supposed to be independent. Note that SBM was originally described in a more general setting, allowing any discrete relational data. However, as explained previously, we concentrate in the following on binary edges only. 

\newpage
Thus, when considering an undirected graph without self loops, this leads to 
\begin{displaymath} 
  p(\Z|    \balpha)   =    \prod_{i=1}^{N}   \mathcal{M}
  (\Z_{i};\: 1, \balpha) = \prod_{i=1}^{N} \prod_{q=1}^{Q}
  \alpha_{q}^{Z_{iq}}, 
\end{displaymath}
and 
\begin{equation*}
  \begin{aligned}
  p(\X | \Z, \bpi) &= \prod_{i < j} p(X_{ij} |
  \Z_{i}, \Z_{j}, \bpi)\\
 & = \prod_{i < j} \prod_{q,
    l}\mathcal{B}(X_{ij}| \pi_{ql})^{Z_{iq}Z_{jl}} \\
 &= \prod_{i < j} \prod_{q,l} \left(\pi_{ql}^{X_{ij}}(1-\pi_{ql})^{1-X_{ij}}\right)^{Z_{iq}Z_{jl}}.
  \end{aligned}
\end{equation*}
In the case of a directed graph, the products over
$i  < j$  must be  replaced by  products over  $i \neq  j$.  The edges
$X_{ii}$ must also be taken into account if the graph contains self-loops.

Note that SBM is related to the infinite block model of \cite{techreport:Kemp:04} although the number $Q$ of classes is fixed. Moreover, contrary to the mixed membership  stochastic block
model of \cite{Article:Airoldi:2008} which captures partial membership and allows each
vertex  to  have a
distribution over a set of classes, SBM assumes
that each vertex of a network belongs to a single class.

The identifiability of SBM was studied by \cite{Article:Matias:2009}, 
who showed that the model is generically identifiable up to a permutation of the classes.
 In other words, except in a set of parameters which has a null Lebesgue's measure,
 two parameters imply the same random graph model if and only if they differ
 only by the ordering of the classes.

\subsection{A Bayesian Stochastic Block Model}
\label{Section:bayesian}

SBM  can be described in a full Bayesian
framework where it  can be considered as a generalisation of the
affiliation model proposed by \cite{Article:hofman:2008}.  Indeed, the
Bayesian model of \cite{Article:hofman:2008} considers a simple
structure where  vertices of
the  same  class connect  with  probability  $\lambda$ and  with
probability $\epsilon$
otherwise.  Therefore, it can be seen as a constrained SBM where the
diagonal of $\bpi$ is set to $\lambda$ and all the other elements to
$\epsilon$. 

To extend the SBM  frequentist  model, we
first specify some non informative conjugate priors for the model parameters. Since
$p(\Z_{i}| \balpha)$ is a multinomial distribution, we consider a
Dirichlet distribution for the mixing coefficients
\begin{equation}
     p \Big(\balpha| \n^{0} = \{n_{1}^{0}, \dots,
  n_{Q}^{0}\} \Big) = \Dirichlet(\balpha;\: \n^{0}), 
\end{equation}
where  $n_{q}^{0}=1/2,\:\forall q$. This Dirichlet distribution corresponds to a non-informative Jeffreys
prior distribution which is known to be proper
\citep{Proceedings:Jeffreys:1946}. It is also possible to 
 consider a uniform distribution on  the $Q-1$ dimensional
simplex by fixing $n_{q}^{0}= 1 , \forall q$. 

Since $p(X_{ij}|\Z_{i}, \Z_{j}, \bpi)$ is a Bernoulli
distribution, we  use independent $\Beta$ priors to  model the  connectivity matrix
$\bpi$
\begin{equation}
  p  \Big(\bpi|   \boeta^{0}  =  (\eta_{ql}^{0}),
  \bozeta^{0} = (\zeta_{ql}^{0}) \Big) = 
  \prod_{q \leq l}
  \Beta(\pi_{ql};\: \eta_{ql}^{0}, \zeta_{ql}^{0}),  
 \end{equation}
with $\eta_{ql}^{0}  =  \zeta_{ql}^{0}  =  1/2,\forall
q$. This corresponds to a product of non-informative Jeffreys prior distributions. Note that if the
graph is directed, the products over $q \leq l$, must be
replaced by products over $q,l$ since $\bpi$ is no longer symmetric.

\begin{figure}[here] 
  \centering 
  \includegraphics[width=6cm, height=6cm]{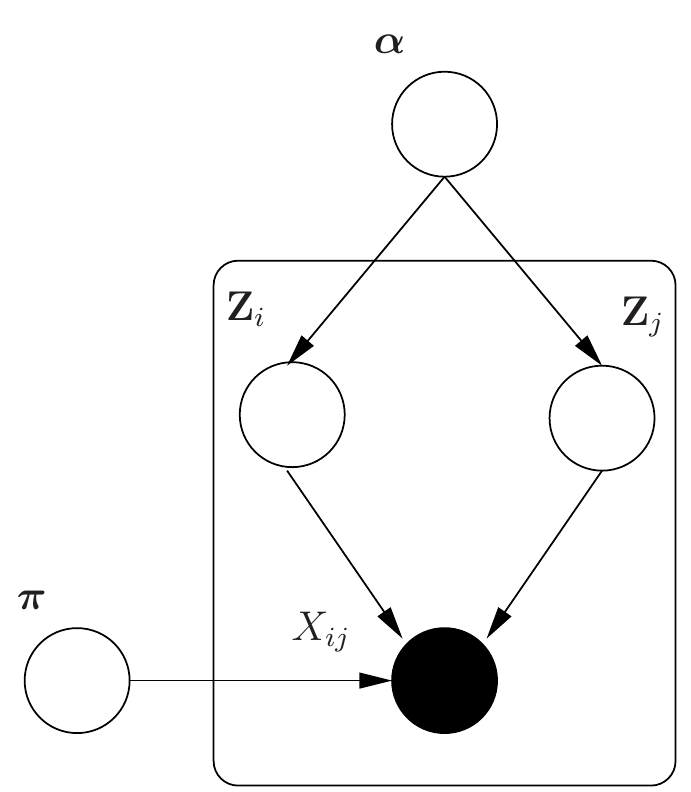}
  \caption{Directed  acyclic graph representing  the Bayesian  view of
  the stochastic block  model. Nodes  represent random variables, which are shaded when they are observed and edges represent conditional dependencies.}
   \label{Fig:probBayes} 
\end{figure}

Thus,  the  model  parameters   are  now  seen  as  random  variables
(see Figure \ref{Fig:probBayes}) whose distributions depend on the hyperparameters  $\n^{0}$,
$\boeta^{0}$, and $\bozeta^{0}$. 
In the following, since these hyperparameters are fixed
and in  order to  keep the  notations simple, they  will not  be shown
explicitly in the conditional distributions.

\captionsetup[figure]{position=bottom, labelfont=bf, labelsep=quad}


\section{Estimation} \label{Section:estimation}

In this section, we first describe the variational EM algorithm used by
\cite{Article:Daudin:2008} to jointly estimate SBM model
parameters and cluster the vertices of a network. We then propose a
new variational Bayes EM algorithm for SBM which approximates the
full posterior distribution of the model parameters and latent variables,
given the observed data $\X$. This procedure relies on a lower bound which
will be later used, in Section \ref{Section:modelSelection}, as a non
asymptotic approximation of the marginal log-likelihood $\ln p(\X)$.

\subsection{Variational Approach} \label{subsection:ML}

The    likelihood    $p(\X|    \balpha,
\bpi)$ of the observed data $\X$ can be obtained through the
marginalization $p(\X| \balpha,
\bpi) = \sum_{\Z}p(\X, \Z| \balpha,
\bpi)$.   This summation  involves $Q^{N}$  terms  and quickly
becomes intractable.  To tackle such problem, the well known EM algorithm \citep{Article:Dempster:1977, mac-kri} has been applied with
success on a large variety of mixture models.  This two stage
estimation approach
\citep{Article:Hathaway:1986,Proc:neal:1998}  can be
described in a variational inference framework. Thus, given a
distribution $q(\Z)$ over the latent variables, 
the
$\log$-likelihood of the observed data is decomposed into two terms
\begin{equation} \label{Eq:decompFreq}
  \ln   p(\X|\balpha,  \bpi)   =  \mathcal{L}\Big(q(.);
  \:\balpha,      \bpi\Big)      +      \KL\Big(q(.)\:||\:
  p(.| \X, \balpha, \bpi)\Big), 
\end{equation}
where 
\begin{equation} \label{Eq:lowFreq}
  \mathcal{L}\Big(q(.);\:     \balpha,     \bpi\Big)     =
  \sum_{\Z} q(\Z)  \ln \{\frac{p(\X, \Z|
    \balpha, \bpi)}{q(\Z)}\}, 
\end{equation}
and 
\begin{equation} \label{Eq:KLFreq}
  \KL\Big(q(.)\:||\:
  p(.|\X, \balpha, \bpi)\Big) = -
  \sum_{\Z} q(\Z) \ln \{\frac{p(\Z| \X,
    \balpha, \bpi)}{q(\Z)}\}.
\end{equation}
In (\ref{Eq:decompFreq}) and (\ref{Eq:KLFreq}), KL denotes the
Kullback-Leibler divergence between the distribution $q(\Z)$ and the
distribution $p(\Z|\X, \balpha,
\bpi)$. Suppose that the current value of the model parameters is
$(\balpha^{old},\bpi^{old})$. During the E-step, the lower bound $ \mathcal{L}\Big(q(.);\:
\balpha^{old},     \bpi^{old}\Big)$ is maximized with respect to
$q(\Z)$ while holding the model parameters fixed. The solution to this
optimization step occurs when the KL divergence vanishes, that is when
$q(\Z)$ is equal to $p(\Z|\X, \balpha^{old},     \bpi^{old})$. The
lower bound is then equal to the log-likelihood of the observed data.
In the M-step, the distribution $q(\Z)$ is held fixed and the lower bound is maximized with respect to the model parameters to give
$(\balpha^{new}, \bpi^{new})$. This causes the log-likelihood to increase.

Unfortunately, when considering SBM,  $p(\Z|\X, \balpha, \bpi)$ 
is not tractable and variational approximations are
required. It  can  be  easily  verified that  minimizing  (\ref{Eq:KLFreq}) with
respect to $q(\Z)$ is
equivalent  to  maximizing   the  lower  bound  (\ref{Eq:lowFreq})  of
(\ref{Eq:decompFreq}) with respect to $q(\Z)$. To obtain a tractable
algorithm, \cite{Article:Daudin:2008} assumed that the distribution
$q(\Z)$ can be factorized such that
\begin{displaymath}
  q(\Z)  = \prod_{i=1}^{N}  q(\Z_{i})  = \prod_{i=1}^{N}
  \mathcal{M}(\Z_{i};\:1,  \btau_{i}),
\end{displaymath}
where $\tau_{iq}$ is a variational parameter denoting the probability of node $i$ to belong to
class $q$.
This gives rise to a so-called variational EM procedure.  During the variational E-step, the model parameters are fixed and, by
maximizing (\ref{Eq:lowFreq}) with respect to $q(\Z)$, the algorithm looks for an
approximation of the conditional distribution of the latent
variables. Conversely, during the variational M-step, the
approximation $q(\Z)$ is fixed and the lower bound is maximized with
respect to the model parameters. This procedure is repeated until
convergence and was proposed by \cite{Article:Daudin:2008} for the SBM model.

\subsection{Variational Bayes EM} \label{subsubsection:vainfer}
In the context of mixture models, the conditional distribution
$p(\Z|\X, \balpha, \bpi)$ can generally be computed  and therefore
Bayesian inference strategies focus on estimating the posterior
distribution $p(\balpha, \bpi|\X)$. The distribution $p(\Z, \balpha,
\bpi|\X)$ is then simply given by a byproduct. However, when
considering SBM, the distribution $p(\Z|\X, \balpha,
\bpi)$ is intractable and so we propose to approximate the full distribution $p(\Z, \balpha,
\bpi|\X)$. We follow  the  work of
\cite{Proceedings:Aatias, Proceedings:Corduneanu:2001, Article:Svensen:2004} on Bayesian mixture modelling and
Bayesian model selection. 
Thus, the marginal log-likelihood, also called integrated
\emph{observed-data} log-likelihood, can be decomposed into two terms
\begin{equation} \label{Eq:decompBayes}
  \ln p(\X) = \mathcal{L}\Big(q(.)\Big) + \KL\Big(q(.)\:||\:
  p(.|\X)\Big), 
\end{equation}
where
\begin{equation} \label{Eq:lowBayes}
  \mathcal{L}\Big(q(.)\Big) =  \sum_{\Z} \int \int q(\Z,
  \balpha, \bpi) \ln
  \{\frac{p(\X,         \Z,         \balpha,
    \bpi)}{q(\Z, \balpha, \bpi)}\} d \balpha d \bpi,
\end{equation}
and
\begin{multline} \label{Eq:KLBayes}
\KL\Big(q(.)\:||\:
p(.|\X)\Big) \\= -
\sum_{\Z} \int \int q(\Z, \balpha, \bpi) \ln \{\frac{p(\Z,
  \balpha,                \bpi               |
  \X)}{q(\Z, \balpha, \bpi)}\} d \balpha d \bpi .
\end{multline}
Again,    as    for    the    variational   EM    approach    (Section
\ref{subsection:ML}),  minimizing (\ref{Eq:KLBayes}) with respect to  $q(\Z, \balpha, \bpi)$ is  equivalent to
maximizing     the      lower     bound     (\ref{Eq:lowBayes})     of
(\ref{Eq:decompBayes}) with respect to $q(\Z, \balpha, \bpi)$.  However,  we  now  have  a  full  variational
optimization problem  since the model parameters  are random variables
and    we are  looking   for    an  approximation   $q(\Z,
\balpha,  \bpi)$ of $p(\Z, \balpha,
\bpi|\X)$. To obtain a tractable algorithm, we assume that the
distribution $q(\Z,
\balpha,  \bpi)$ can be factorized such that 
\begin{displaymath} \label{Eq:MixNetFactor}
  q(\Z, \balpha, \bpi) =  q(\balpha) q(\bpi) q(\Z) =  q(\balpha) q(\bpi)\prod_{i=1}^{N} q(\Z_{i}).
\end{displaymath}
In the following, we use a variational Bayes EM algorithm. We call
variational  Bayes  E-step,  the  optimization  of  each  distribution
$q(\Z_{i})$ and  variational Bayes M-step,  the approximations
of the remaining distributions $q(\balpha)$ and $q(\bpi)$. 
All the optimization equations, the lower bound, as well as proofs are
given in the appendix. 

We first initialize a matrix $\btau^{old}$ with a
hierarchical algorithm based on  the
classical Ward distance. The distance between vertices which is
considered is simply the Euclidean distance $d(i, j) = \sum_{k=1}^{N}
(X_{ik} - X_{jk})^{2}$ which takes the number of discordances between
$i$ and $j$ into account. Given a number of classes $Q$, each vertex
is assigned (hard assignment) to its nearest group. Second, the algorithm uses (\ref{Equation:dirichlet}) and (\ref{Equation:Beta}) to estimate
the variational distributions over the model parameters $\balpha$ as
well as $\bpi$. Finally,  the variational distribution
over the latent variables is estimated using
(\ref{Equation:multinomial}). The algorithm cycles though
the E and M steps until the absolute distance between two successive
values of the lower bound (\ref{Equation:lowerbound}) is smaller than a
threshold $eps$. In the experiment section, we set $eps=1e-6$.
 In practice, smaller values slow the convergence of the algorithm and
 do not lead to better estimates.

The computational costs of the frequentist
approach  of  \cite{Article:Daudin:2008}  and  our  variational  Bayes
algorithm are both equal to $\mathrm{O}(Q^{2}N^{2})$. Analyzing a sparse network  takes about a second for $N=200$ nodes and about a minute for $N=1000$.

\section{Model selection} \label{Section:modelSelection}

So far,  we have seen that  the variational Bayes  EM algorithm leads
to an approximation of the posterior distribution of all the model parameters and latent
variables, given the observed data. However, the problem of
estimating the number $Q$ of classes in the mixture has not been
addressed yet. 
Given a set of values of $Q$, we aim at selecting
$Q^{*}$ which maximizes the  marginal log-likelihood $\ln p(\X|
Q)$, also called integrated
\emph{observed-data} log-likelihood. The marginal likelihood is known to focus on density
estimation view and is expected to provide a consistent estimation of
the distribution of the data \citep{article:Biernacki:2010}. Unfortunately, this quantity is not tractable, since for each value of $Q$, it involves
integrating  over  all  the  model parameters and latent variables
\begin{displaymath}
\ln p(\X|Q) = \ln \left\{\sum_{\Z} \int \int p(\X, \Z, \balpha,
\bpi|Q)\:d\balpha d\bpi\right\}.
\end{displaymath}
To tackle this issue, we propose to replace the marginal log-likelihood
with its variational Bayes approximation. Thus, given a value of
$Q$, the algorithm introduced in Section \ref{subsubsection:vainfer} is used to maximize the lower bound
(\ref{Eq:lowBayes}) with respect to $q(.)$. We recall that this
maximization implies a minimization of  the KL divergence (\ref{Eq:KLBayes}) between $q(.)$ and the unknown
posterior distribution. After convergence of the algorithm,
according to (\ref{Eq:decompBayes}), if the KL divergence is small, then the lower bound
$\mathcal{L}\Big(q(.)\Big)$ approximates the marginal log-likelihood.
Obviously, this assumption  can not be verified in
practice since (\ref{Eq:KLBayes}) can not be computed analytically.
Moreover, we emphasize
that there is no solid reason to believe that the KL divergence is close
to zero and does not depend on the model complexity. Nevertheless, in
order to obtain a tractable model selection criterion we rely on this approximation.
After convergence of the algorithm, the lower
bound takes a simple form and leads to a new criterion for SBM that we
call ILvb
\begin{multline} 
    IL_{vb} =  \ln\{\frac{\Gamma(\sum_{q=1}^{Q}n_{q}^{0})
      \prod_{q=1}^{Q}\Gamma(n_{q})}{\Gamma(\sum_{q=1}^{Q}n_{q})\prod_{q=1}^{Q}\Gamma(n_{q}^{0})}\}
  \\  +   \sum_{q   \leq   l}^{Q}   \ln   \{\frac{\Gamma(\eta_{ql}^{0}   +
      \zeta_{ql}^{0})\Gamma(\eta_{ql})\Gamma(\zeta_{ql})}{\Gamma(\eta_{ql}              +
      \zeta_{ql})\Gamma(\eta_{ql}^{0})\Gamma(\zeta_{ql}^{0})}\} 
    - \sum_{i=1}^{N}\sum_{q=1}^{Q} \tau_{iq} \ln \tau_{iq},
\end{multline}
where $\tau_{iq}$ is the estimated probability of vertex $i$ to belong
to class $q$ and $(n_{q})_{q}$, $(\eta_{ql})_{ql}$, $(\zeta_{ql})_{ql}$ are parameters
given in the appendix. The gamma function is denoted by $\Gamma(\cdot)$.
Contrary to the criterion proposed by \cite{Article:Daudin:2008}, ILvb
does not rely on an asymptotic approximation, sometimes called
BIC-like approximation. In practice, given a network, the variational Bayes EM algorithm is run for
the different values of $Q$ considered and $Q^{*}$ is chosen such that
ILvb is maximized.

\section{Experiments} 
\label{Section:experiments}

We present some results of the experiments we carried out 
to assess  the criterion we proposed in Section
\ref{Section:modelSelection}. Throughout our experiments,
 we chose to compare our approach to the work of
\cite{Article:Daudin:2008}   and  \cite{Article:hofman:2008}.  Indeed,
contrary to many other \emph{model based techniques}, the corresponding algorithms can
analyze networks  with hundred of nodes in  a reasonable
amount of time (a few minutes on a dual core). We recall
that   \cite{Article:Daudin:2008}  proposed   a  frequentist  maximum
likelihood approach (see Section \ref{subsection:ML}) for SBM as well as an ICL criterion. On the other
hand,  \cite{Article:hofman:2008}  presented   a  model  for
community structure detection  and a Bayesian criterion
that we will denote VBMOD. Thus, by using both synthetic data and the metabolic
network of bacteria \emph{Escherichia coli}, our aim is twofold. 
First, we illustrate the overall capacity of SBM to retrieve
interesting  structures in a  large variety  of networks. Second, we concentrate on comparing the two
criteria ICL and ILvb developed for SBM.

\subsection{Comparison of the criteria}

In these experiments, we consider two types of networks.  In Section
\ref{subsubsection:affNetworks}, we
generate affiliation networks, made of community structures, using the
generative model of \cite{Article:hofman:2008}. Therefore, vertices of
the  same  class connect  with  probability  $\lambda$ and  with
probability $\epsilon$
otherwise.  This  corresponds to a constrained SBM  where the diagonal
of  the connectivity  matrix is  set to  $\lambda$ and  all  the other
elements to $\epsilon$
\begin{displaymath}
  \bpi = \begin{pmatrix}
    \lambda & \epsilon & \dots & \epsilon\\
    \epsilon & \lambda &  & \vdots\\
     \vdots & & \ddots & \epsilon\\
     \epsilon & \dots & \epsilon & \lambda
  \end{pmatrix}.
\end{displaymath}

In Section \ref{subsubsection:affAndHubNetworks}, we then draw networks with more complex topologies, made of both community structures
and a class of hubs.  The
corresponding model is given by the connectivity matrix
\begin{displaymath}
  \bpi = \begin{pmatrix}
    \lambda & \epsilon & \dots & \epsilon & \lambda\\
    \epsilon & \lambda & & & \vdots\\
     \vdots & & \ddots & & \vdots\\
     \lambda & \dots & \dots & \dots & \lambda
  \end{pmatrix},
\end{displaymath}
where hubs  connect with probability $\lambda$ to  any vertices in
the network.

 Following \cite{Article:Mariadassou:2010} who showed that ICL tends to
underestimate the number of classes in the case of small graphs, we
consider  networks  with  only $N=50$  vertices  to analyze  the
robustness of our criterion.  We set $(\lambda=0.9,\:\epsilon=0.1)$
and for each value of $Q_{True}$ in the set $\{3,
\dots, 7\}$, we then generate 100 networks with classes mixed in the same proportions $\alpha_{1} = \dots
=  \alpha_{Q_{True}} = 1/Q_{True}$. 

 In order to
estimate the  number of classes  in the latent structures,  we applied
the methods of \cite{Article:hofman:2008}, \cite{Article:Daudin:2008},
and our algorithm (Section \ref{subsubsection:vainfer}) on each network, for various numbers of classes
$Q \in \{1, \dots, 7\}$.  Note that, we choose $n_{q}^{0}
= 1/2,\: \forall q \in \{1, \dots, Q\}$ for the Dirichlet prior and $\eta_{ql}^{0} =
\zeta_{ql}^{0} = 1/2,\: \forall (q,  l) \in \{1, \dots, Q\}^{2}$ for the
Beta priors.  We recall that such distributions  correspond to non
informative prior distributions. Like
any optimization technique, the clustering methods we consider depend on the
initialization. Thus, for each simulated network and each number of classes
$Q$, we use five different
initializations  of $\boldsymbol{\tau}$.  Finally,  we select the best
learnt models for which the  corresponding criteria VBMOD, ICL, or ILvb
were maximized.

Before comparing  ICL and ILvb, it is
crucial to recall that these two criteria were not conceived for the same
purpose. ICL approximates the integrated \emph{complete-data}
likelihood and is known to focus on cluster analysis view since it
favors well separated clusters. It realizes a compromise between the
estimation of the data density and the evidence of data
partitioning. Conversely, ILvb approximates the
marginal likelihood which is known to focus on density
estimation only. In the following experiments, since networks are
generated using SBM, and because we evaluate the criteria through
their
capacity to retrieve the true number of classes, ILvb is expected to
lead to better results. However, in other situations (which are not
considered in this paper), where the focus would be on the clustering of
vertices, ICL might be of possible interest.

\subsubsection{Affiliation Networks}
\label{subsubsection:affNetworks}

In  Table \ref{table:aff0.9-0.1},  we observe  that  VBMOD outperforms
both ICL and ILvb.  For instance, when $Q_{True}=5$, VBMOD correctly
estimates the number of classes of the 100 generated networks, while ICL and
ILvb have  respectively a  percentage of accuracy  of 77 and  99. These
differences increase  when $Q_{True}=6$ and  $Q_{True}=7$. Indeed, the
higher  $Q_{True}$ is,  the  less vertices  the  classes contain,  and
therefore, the more difficult it is to retrieve and distinguish the community structures.  Thus, when $Q_{True}=7$, each
class  only contains on  average  $Q_{True}/N\approx 7.1$  vertices. VBMOD appears to  be a very stable  criterion for
community structure detection. It has a percentage of accuracy of 84 while ICL
never  estimates the  true  number of  classes. 

All  the  affiliation  networks  were  generated using  the  model  of
\cite{Article:hofman:2008} which explains the results of VBMOD presented
above. Indeed,  the corresponding  model for community structure detection only
estimates   the  parameters   $\lambda$  and   $\epsilon$   whereas  the
frequentist and Bayesian approaches for SBM look for a full $Q \times Q$
matrix  $\bpi$  of  connection  probabilities.  They  are  capable  of
handling networks  with complex topologies, as shown  in the following
section, but they might miss some structures if the number of vertices
is too limited.

We observe that ILvb leads to a  better estimates of the true number of classes in
networks than ICL. Thus, when $Q_{True} = 5$ and $Q_{True} = 6$, ILvb estimates correctly
the number of classes of 99 and 73 networks while ICL has respectively
a percentage of accuracy of 77 and 12.

\captionsetup[table]{position=bottom, labelfont=bf, labelsep=quad}
\begin{table}[h] 
  \caption{Confusion matrices for VBMOD, ICL and ILvb.  $\lambda = 0.9$, $\epsilon = 0.1$ and
  $Q_{True} \in \{3, \dots, 7\}$. Affiliation networks.}
  \centering
  \subfloat[$Q_{True} \backslash Q_{VBMOD}$]{
    \begin{tabular}{c| c c c c c c }
       & 2 & 3 & 4 & 5 & 6 & 7\\
      \hline
      3 & 0 & 100 & 0 & 0 & 0 & 0\\
      4 & 0 & 0 & 100 & 0 & 0 & 0\\
      5 & 0 & 0 & 0 & $\mathbf{100}$ & 0 & 0\\
      6 & 0 & 0 & 0 & 0 & $\mathbf{97}$ & 3\\
      7 & 0 & 0 & 0 & 2 & 14 & $\mathbf{84}$\\
    \end{tabular}
  }
  \\
  \subfloat[$Q_{True} \backslash Q_{ICL}$] {
    \begin{tabular}{c| c c c c c c}
      & 2 & 3 & 4 & 5 & 6 & 7\\
      \hline 
      3 & 0 & 100 & 0 & 0 & 0 & 0\\
      4 & 0 & 0 & 100 & 0 & 0 & 0\\
      5 & 0 & 0 & 23 & $\mathbf{77}$ & 0 & 0\\
      6 & 0 & 1 & 28 & 59 & $\mathbf{12}$ & 0\\
      7 & 0 & 8 & 49 & 42 & 1 & $\mathbf{0}$\\
    \end{tabular}
  }
  \\
  \subfloat[$Q_{True} \backslash Q_{ILvb}$] {
    \begin{tabular}{c| c c c c c c}
      & 2 & 3 & 4 & 5 & 6 & 7\\
      \hline 
      3 & 0 & 100 & 0 & 0 & 0 & 0\\
      4 & 0 & 0 & 100 & 0 & 0 & 0\\
      5 & 0 & 0 & 0 & $\mathbf{99}$ & 1 & 0\\
      6 & 0 & 0 & 4 & 23 & $\mathbf{73}$ & 0\\
      7 & 0 & 2 & 14 & 44 & 27 & $\mathbf{13}$\\
    \end{tabular}
  }
  \label{table:aff0.9-0.1}
\end{table}

\subsubsection{Networks with Community Structures and Hubs}
\label{subsubsection:affAndHubNetworks}

Table \ref{table:0.9-0.1} displays the results of the experiments on
networks exhibiting community structures and hubs.  The presence of
hubs is a central property of so-called real real networks
\citep{Article:Albert:2002}.

This slightly more complex and more
realistic situation does heavily perturb the estimation of VBMOD.  Most of
the time, VBMOD fails to detect the class of hub and henceforth
underestimates the number of classes. For example, when $Q_{True}=3$
or $Q_{True}=4$,  VBMOD always misses a class. When the number of true
classes grows over four,   VBMOD's behaviour becomes more variable but keep 
the same heavy tendency to underestimate. 

In this context, ICL and ILvb  behaves more consistently than VBMOD.
 When $Q_{True}$ is less or equal  than four both strategies are comparable.
But when the number of true classes increases, the performance of ICL
dramatically deteriorates, whereas ILvb remains more stable. 

In the context of small graph, when the focus is on the estimation of
the data density, ILvb clearly provides a more reliable
estimation of the number of class than ICL. It also shows better
performances that VBMOD when networks are made of classes which are
not communities.

\captionsetup[table]{position=bottom, labelfont=bf, labelsep=quad}
\begin{table}[h] 
  \caption{Confusion matrices for VBMOD, ICL and ILvb.  $\lambda = 0.9$, $\epsilon = 0.1$ and
  $Q_{True} \in \{3, \dots, 7\}$. Affiliation networks and a class of hubs.}
  \centering
  \subfloat[$Q_{True} \backslash Q_{VBMOD}$]{
    \begin{tabular}{c| c c c c c c }
      & 2 & 3 & 4 & 5 & 6 & 7\\
      \hline
      3 & 95 & $\mathbf{0}$ & 3 & 0 & 0 & 2\\
      4 & 1 & 95 & $\mathbf{4}$ & 0 & 0 & 0\\
      5 & 0 & 0 & 94 & $\mathbf{6}$ & 0 & 0\\
      6 & 0 & 0 & 1 & 83 & $\mathbf{16}$ & 0\\
      7 & 0 & 0 & 2 & 15 & 78 & $\mathbf{5}$\\
    \end{tabular}
  }
  \\
  \subfloat[$Q_{True} \backslash Q_{ICL}$] {
    \begin{tabular}{c| c c c c c c}
      & 2 & 3 & 4 & 5 & 6 & 7\\
      \hline 
      3 & 0 & $\mathbf{100}$ & 0 & 0 & 0 & 0\\
      4 & 0 & 0 & $\mathbf{100}$ & 0 & 0 & 0\\
      5 & 0 & 0 & 12 & $\mathbf{88}$ & 0 & 0\\
      6 & 0 & 0 & 19 & 59 & $\mathbf{22}$ & 0\\
      7 & 0 & 3 & 29 & 56 & 12 & $\mathbf{0}$\\
    \end{tabular}
  }
  \\
  \subfloat[$Q_{True} \backslash Q_{ILvb}$] {
    \begin{tabular}{c|c c c c c c}
      & 2 & 3 & 4 & 5 & 6 & 7\\
      \hline 
      3 & 0 & $\mathbf{100}$ & 0 & 0 & 0 & 0\\
      4 & 0 & 0 & $\mathbf{100}$ & 0 & 0 & 0\\
      5 & 0 & 0 & 2 & $\mathbf{98}$ & 0 & 0\\
      6 & 0 & 0 & 1 & 29 & $\mathbf{70}$ & 0\\
      7 & 0 & 0 & 3 & 34 & 45 & $\mathbf{18}$\\
    \end{tabular}
  }
  \label{table:0.9-0.1}
\end{table}

\newpage
\subsection{The metabolic network of \emph{Escherichia coli}}
We  apply the  methodology described  in this  paper to  the metabolic
network of bacteria \emph{Escherichia coli}
\citep{Article:Lacroix:2006} which was analyzed by
\cite{Article:Daudin:2008} using SBM. In this network, there
are 605 vertices which represent chemical reactions and a total number of 1782 edges. Two reactions are connected
if a compound  produced by the first  one is a part of  the second one
(or vice-versa).  As in the previous section, we consider non informative
priors: we fixed $n_{q}^{0}
= 1/2,\: \forall q \in \{1, \dots, Q\}$ for the Dirichlet prior and $\eta_{ql}^{0} =
\zeta_{ql}^{0} = 1/2,\: \forall (q,  l) \in \{1, \dots, Q\}^{2}$ for the
Beta priors.

\captionsetup[figure]{position=bottom, labelfont=bf, labelsep=quad}
\begin{figure}[h] 
  \centering 
  \includegraphics[width=8cm, height=8cm]{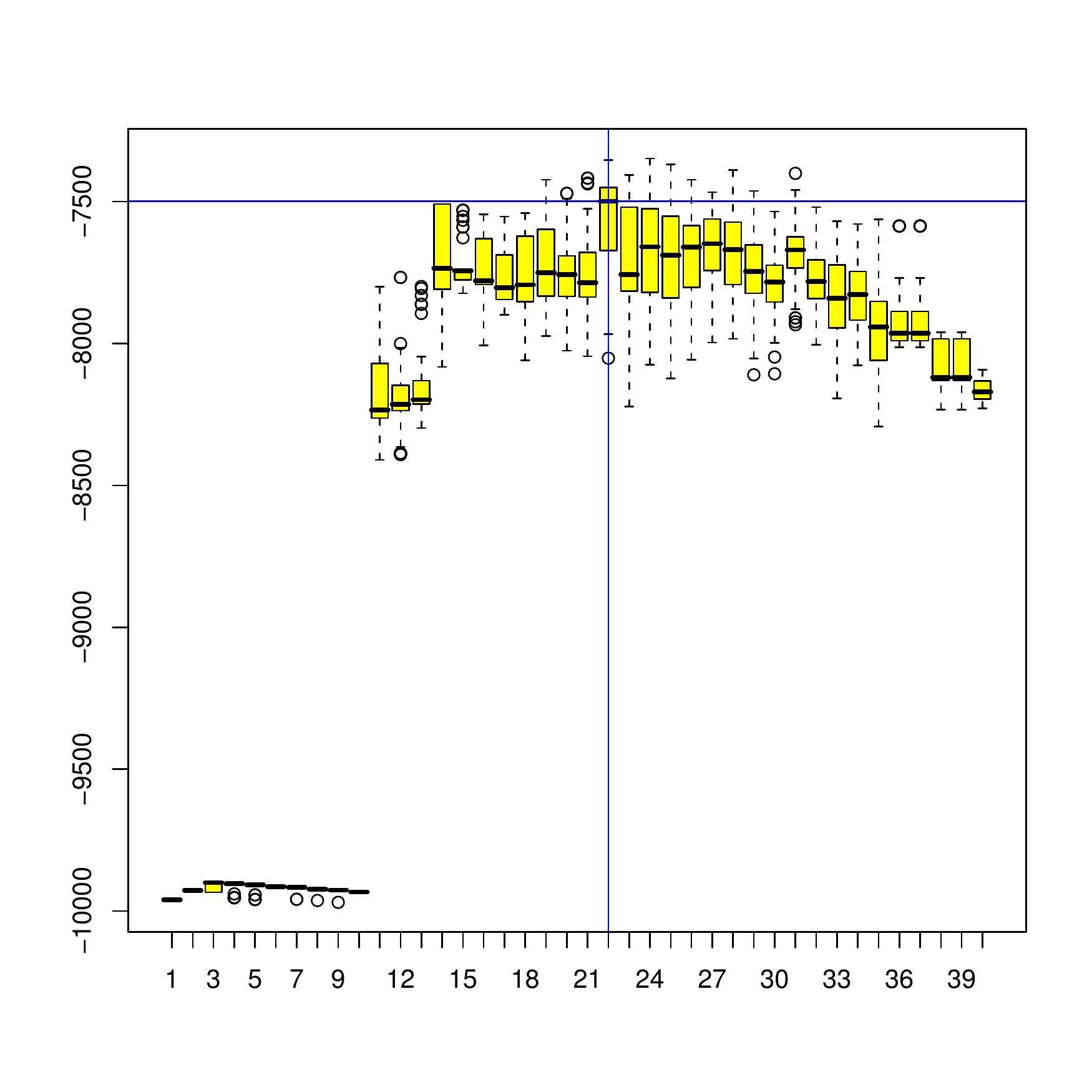}
  \caption{Boxplot representation (over 60 experiments) of ILvb for $Q
    \in \{1, \dots, 40\}$. The maximum is reached at $Q_{IL_{vb}} = 22$.}
   \label{Figure:vb60rep} 
\end{figure}

Thus,  for  $Q  \in  \{1,   \dots,  40\}$,  we  apply  the  methods  of
\cite{Article:hofman:2008}  as well as   our
approach on  this network. We compute the  corresponding criteria and
we repeat such procedure
60 times, for different initializations of $\boldsymbol{\tau}$.  Indeed, to speed up the initialization, we first run a kmeans algorithm with $40$ classes and random initial centers. We then use the corresponding partitions as inputs of the hierarchical algorithm described in Section \ref{subsubsection:vainfer}. The results for ILvb are presented as boxplots in Figure
\ref{Figure:vb60rep}. The  criterion finds  its maximum for  $Q_{ILvb} =
22$ classes, while \cite{Article:Daudin:2008} found $Q_{ICL} = 21$. Thus, for this
particular large data set, both ILvb and ICL lead to almost the same estimates of
the  number  of  latent  classes.

\captionsetup[figure]{position=bottom, labelfont=bf, labelsep=quad}
\begin{figure}[h]
  \centering 
  \includegraphics[width=8cm, height=8cm]{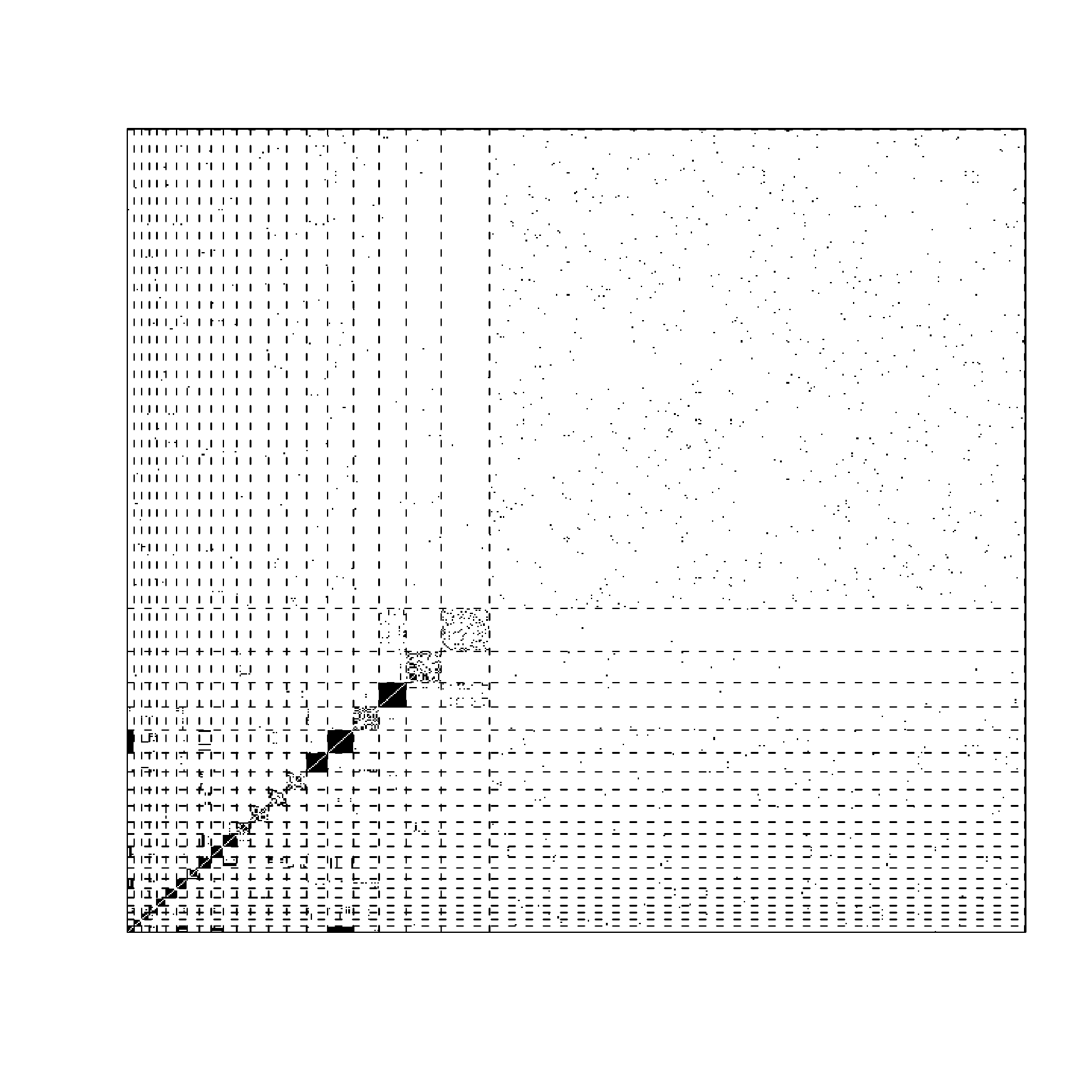}
  \caption{Dot  plot  representation of  the  metabolic network  after
    classification of the vertices into $Q_{VB} = 22$ classes. The
    x-axis and y-axis correspond to  the list of vertices in the network, from $1$
    to $605$. Edges between pairs of vertices are represented by shaded dots.} 
   \label{Figure:dotRepresentation} 
\end{figure}

We  also compared the  learnt partitions  in the  Bayesian and  in the
frequentist approach.  Figure \ref{Figure:dotRepresentation} is  a dot
plot representation of the  metabolic network after having applied the
Bayesian algorithm for $Q_{VB} = 22$. Each vertex $i$ is classified
into the class for which $\tau_{iq}$ is maximal (Maximum A Posteriori
estimate).  We observed very similar patterns in the frequentist
approach.  
Among the classes, eight of them are cliques $\pi_{qq}=1$ and six have within
probability connectivity greater than 0.5. As shown by
\cite{Article:Daudin:2008},  these cliques or pseudo-cliques gather
reactions involving a same compound.  Thus, chorismate, pyruvate,
L-aspartate, L-glutamate, D-glyceraldehyde-3-phosphate and ATP are
all responsible for cliques. Moreover, as observed in
\cite{Article:Daudin:2008}, since the connection probability between class 1
and 17 is 1, they correspond to a single clique which is associated to
pyruvate. However that clique  is
split into  two sub-cliques because of  their different connectivities
with reactions of classes 7 and 10. 
Since the approach of
\cite{Article:hofman:2008}   only  looks  for   community  structures,
it  can not  retrieve such  complex  topologies, as  shown in  Section
\ref{subsubsection:affAndHubNetworks}, and many classes such as
class 1 and 17 were merged. We found $Q_{VBMOD}=14$.

\section{Conclusion}
\label{Section:conclusion}

In this paper, we showed how the Stochastic Block Model (SBM) could be described in a full Bayesian
framework. We introduced some non informative conjugate priors over the
model parameters and we described a variational Bayes EM
algorithm which approximates the
posterior distribution of all  the latent variables and model
parameters, given the observed data.
Using this framework, we derived a non asymptotic model selection
criterion, so-called ILvb, which approximates the marginal likelihood. By considering
networks generated using SBM, we showed that ILvb focus on
the estimation of the data density and provides a relevant estimation of
the number of latent classes. We also illustrated the capacity of SBM to  retrieve  interesting  structures  in  a  large  variety  of
networks, contrary to algorithms looking for community structures
only. In future work, we will investigate approximate Bayesian
computation methods for model selection. These simulation techniques
seem particularly promising for the analysis of SBM where the likelihood of
the observed data is intractable.


\newpage
\onecolumn
\appendix
\section*{Appendix}

\section{Approximation of   $q(\Z_{i})$ the conditional distribution of the latent variables}

  The optimal approximation at vertex $i$ is
  \begin{equation} \label{Equation:multinomial}
    q(\Z_{i})      =      \mathcal{M}(\Z_{i};\:     1,
    \boldsymbol{\tau_{i}} = \{\tau_{i1}, \dots, \tau_{iQ}\}),
  \end{equation}
  where $\tau_{iq}$  is the probability (responsability) of
  node $i$ to belong to class $q$. It satisfies the relation
  \begin{equation} \label{Equation:tau}
    \tau_{iq}  \propto  e^{\psi(n_{q})  -
      \psi(\sum_{l=1}^{Q} n_{l})}  \prod_{j \neq i}^{N}\prod_{l=1}^{Q}
    e^{\tau_{jl}\Bigg(\psi(\zeta_{ql})  - \psi(\eta_{ql}  +  \zeta_{ql}) +
      X_{ij} \Big(\psi(\eta_{ql}) - \psi(\zeta_{ql})\Big)\Bigg)}, 
  \end{equation}
  where $\psi(.)$     is    the    \emph{digamma}     function. In order to optimize the distribution $q(\Z)$, we rely on a fixed point algorithm. Thus, given a matrix $\btau^{old}$, the algorithm builds a new matrix $\btau^{new}$ where each rows satisfies (\ref{Equation:tau}). After normalization, it then uses $\btau^{new}$ to build a new matrix and so on. The algorithm stops when $\sum_{i=1}^{N}\sum_{q=1}^{Q}|\tau_{iq}^{old}-\tau_{iq}^{new}|< eps$. In the experiment section, we set $eps = 1e-6$.

{\bf Proof:}
  According  to variational Bayes,  the optimal
  distribution $q(\Z_{i})$ is given by
  \begin{equation} \label{Equation:Z}
    \begin{aligned}
      \ln q(\Z_{i}) & = \E_{\Z^{\backslash i},
        \balpha,     \bpi}[\ln    p(\X,
      \Z, \balpha, \bpi)] + \cst \\
      &=     \E_{\Z^{\backslash     i},     \bpi}[\ln
      p(\X|        \Z,        \boldsymbol{\pi)}]        +
      \E_{\Z^{\backslash     i},     \balpha}    [\ln
      p(\Z| \boldsymbol{\alpha)}] + \cst \\
      &=  \E_{\Z^{\backslash i},  \bpi}[\sum_{i^{'} < j}
      \sum_{q,l}  Z_{i'q} Z_{jl} \Big(X_{i'j}  \ln \pi_{ql}  + (1  - X_{i'j})
      \ln(1 - \pi_{ql})\Big)] \\
      &\qquad \:\:\:\:\: + \E_{\Z^{\backslash i},
        \balpha}  [\sum_{i'=1}^{N}  \sum_{q=1}^{Q} Z_{i'q}  \ln
      \alpha_{q}] + \cst\\
      &= \sum_{q=1}^{Q} Z_{iq} \Bigg( \E_{\alpha_{q}}[\ln \alpha_{q}] + \sum_{j \neq i} ^{N} \sum_{l=1}^{Q}
      \tau_{jl}\Big(X_{ij} \big(\E_{\pi_{ql}}[\ln \pi_{ql}] - \E_{\pi_{ql}}[\ln (1- \pi_{ql})]\big) \\
      &\qquad \:\:\:\:\: + \E_{\pi_{ql}}[\ln (1- \pi_{ql})]\Big) \Bigg) + \cst \\
      &=      \sum_{q=1}^{Q}     Z_{iq}     \Bigg(      \psi(n_{q})     -
      \psi(\sum_{l=1}^{N} n_{l}) + \sum_{j \neq i} ^{N} \sum_{l=1}^{Q}
      \tau_{jl}\Big(X_{ij} \big(\psi(\eta_{ql}) - \psi(\zeta_{ql})\big) \\
      &\qquad   \:\:\:\:\:   +   \psi(\zeta_{ql})  -   \psi(\eta_{ql}   +
      \zeta_{ql})\Big) \Bigg) + \cst, 
    \end{aligned}
  \end{equation} 
  where $\Z^{\backslash i}$ denotes the  class of all nodes except node
  $i$. We have used $\E_{y}[\ln y]= \psi (a) - \psi(a+b)$ when $y \sim \Beta(y;\: a, b)$. Moreover, to simplify the
  calculations, the terms  that do not  depend on $\Z_{i}$
  have  been  absorbed into  the  constant.  Taking  the exponential  of
  (\ref{Equation:Z}) and after normalization, we obtain the multinomial
  distribution (\ref{Equation:multinomial}).

\section{Optimization   of  $q(\balpha)$.}
 The optimization of  the lower  bound with  respect to  $q(\balpha)$  produces   a  distribution  with   the  same
  functional form as the prior $p(\balpha)$
  \begin{equation} \label{Equation:dirichlet}
    q(\balpha) = \Dirichlet(\alpha;\: \mathbf{n}), 
  \end{equation}
  where 
  \begin{equation} \label{Equation:n}
    n_{q} = n_{q}^{0} + \sum_{i=1}^{N} \tau_{iq}.
  \end{equation}

{\bf Proof:}
  According   to    variational   Bayes,   the    optimal   distribution
  $q(\balpha)$ is given by
  \begin{equation} \label{Equation:alpha}
    \begin{aligned}
      \ln q(\balpha) & = \E_{\Z, \bpi}[\ln p(\X, \Z, \balpha, \bpi)] + \cst \\
      & = \E_{\Z}[\ln p(\Z| \balpha)] + \ln p(\balpha) + \cst \\
      & = \sum_{i=1}^{N} \sum_{q=1}^{Q} \tau_{iq} \ln \alpha_{q} + \sum_{q=1}^{Q} (n_{q}^{0} -1)\ln \alpha_{q} + \cst \\
      & = \sum_{q=1}^{Q} \left(n_{q}^{0} -1 + \sum_{i=1}^{N} \tau_{iq}\right) \ln \alpha_{q} + \cst.
    \end{aligned}
  \end{equation}
  Taking   the   exponential   of   (\ref{Equation:alpha})   and   after
  normalization,     we     obtain     the    Dirichlet     distribution
  (\ref{Equation:dirichlet}).

\section{ Optimization   of  $q(\bpi)$.} 
 Again,  the  functional form  of  the  prior $p(\bpi)$  is
  conserved through the variational optimization:
  \begin{equation} \label{Equation:Beta}
    q(\bpi) = \prod_{q \leq l}^{Q} \Beta (\pi_{ql};\: \eta_{ql}, \zeta_{ql}), 
  \end{equation}
  For  $q \neq  l$, the  hyperparameter
  $\eta_{ql}$ is given by
  \begin{equation} \label{Equation:eta1}
    \eta_{ql} = \eta_{ql}^{0} + \sum_{i \neq j}^{N} X_{ij} \tau_{iq} \tau_{jl},
  \end{equation}
  and $\forall q$:
  \begin{equation} \label{Equation:eta2}
    \eta_{qq} = \eta_{qq}^{0} + \sum_{i < j}^{N} X_{ij} \tau_{iq} \tau_{jq}.
  \end{equation}
  Moreover, for $q \neq l$, the hyperparameter $\zeta_{ql}$ is given by
  \begin{equation} \label{Equation:zeta1}
    \zeta_{ql} = \zeta_{ql}^{0} + \sum_{i \neq j}^{N} (1 - X_{ij}) \tau_{iq} \tau_{jl},
  \end{equation}
  and $\forall q$:
  \begin{equation} \label{Equation:zeta2}
    \zeta_{qq} = \zeta_{qq}^{0} + \sum_{i < j}^{N} (1 - X_{ij}) \tau_{iq} \tau_{jq}.
  \end{equation}

{\bf Proof :}
  According   to    variational   Bayes,   the    optimal   distribution
  $q(\bpi)$ is given by
  \begin{equation} \label{Equation:pi}
    \begin{aligned}
      \ln q(\bpi) & = \E_{\Z, \balpha}[\ln p(\X, \Z, \balpha, \bpi)] + \cst \\
      & = \E_{\Z}[\ln p(\X| \Z, \bpi)] + \ln p(\bpi) + \cst \\
      & =  \sum_{i < j}^{N}  \sum_{q,l}^{Q}\tau_{iq} \tau_{jl} \Big(X_{ij}
      \ln \pi_{ql} + (1 - X_{ij})\ln(1 - \pi_{ql})\Big)\\
      & \qquad \:\:\:\:\: + \sum_{q \leq l}^{Q}\Big((\eta_{ql}^{0}-1)\ln \pi_{ql} + (\zeta_{ql}^{0} - 1)\ln(1 - \pi_{ql})\Big) + \cst \\
      & = \sum_{q < l}^{Q}\sum_{i \neq j}^{N} \tau_{iq} \tau_{jl} \Big(X_{ij}
      \ln \pi_{ql} + (1 - X_{ij})\ln(1 - \pi_{ql})\Big) \\
      & \qquad \:\:\:\:\: +  \sum_{q=1}^{Q}\sum_{i < j}^{N} \tau_{iq} \tau_{jq} \Big(X_{ij}
      \ln \pi_{qq} + (1 - X_{ij})\ln(1 - \pi_{qq})\Big)\\
      &  \qquad \:\:\:\:\:  + \sum_{q  \leq l}^{Q}\Big((\eta_{ql}^{0}-1)\ln
      \pi_{ql} + (\zeta_{ql}^{0} - 1)\ln(1 - \pi_{ql})\Big) + \cst \\
      & = \sum_{q < l}^{Q}\Bigg(\Big(\eta_{ql}^{0} - 1 + \sum_{i \neq j}^{N} \tau_{iq} \tau_{jl} X_{ij}\Big)
      \ln   \pi_{ql}   +   \Big(\zeta_{ql}^{0}   -  1   +   \sum_{i   \neq
        j}^{N}\tau_{iq} \tau_{jl}(1  - X_{ij})\Big)\ln(1 - \pi_{ql})\Bigg)
      \\
      & \qquad \:\:\:\:\: + \sum_{q=1}^{Q}\Bigg(\Big(\eta_{qq}^{0} - 1 + \sum_{i < j}^{N} \tau_{iq} \tau_{jq} X_{ij}\Big)
      \ln   \pi_{qq}   +   \Big(\zeta_{qq}^{0}   -  1   +   \sum_{i <
        j}^{N}\tau_{iq} \tau_{jq}(1 - X_{ij})\Big)\ln(1 - \pi_{qq})\Bigg).
    \end{aligned}
  \end{equation}
  Taking   the   exponential   of   (\ref{Equation:pi})   and   after
  normalization, we obtain the product of $\Beta$ distribution
  (\ref{Equation:Beta}).

\section{Lower bound.}

 The  lower bound  takes a  simple form  after the  variational Bayes
  M-step. Indeed, it only depends on the
  posterior probabilities $\tau_{iq}$ as well as the normalizing constants  of the Dirichlet and Beta distributions
  \begin{equation} \label{Equation:lowerbound}
    \mathcal{L}\Big(q(.)\Big)  =  \ln\{\frac{\Gamma(\sum_{q=1}^{Q}n_{q}^{0})
      \prod_{q=1}^{Q}\Gamma(n_{q})}{\Gamma(\sum_{q=1}^{Q}n_{q})\prod_{q=1}^{Q}\Gamma(n_{q}^{0})}\}
  +   \sum_{q   \leq   l}^{Q}   \ln   \{\frac{\Gamma(\eta_{ql}^{0}   +
      \zeta_{ql}^{0})\Gamma(\eta_{ql})\Gamma(\zeta_{ql})}{\Gamma(\eta_{ql}              +
      \zeta_{ql})\Gamma(\eta_{ql}^{0})\Gamma(\zeta_{ql}^{0})}\} 
    - \sum_{i=1}^{N}\sum_{q=1}^{Q} \tau_{iq} \ln \tau_{iq}.
  \end{equation}

{\bf Proof :}
The lower bound is given by 
\begin{equation} 
  \begin{aligned}
    \mathcal{L}\Big(q(.)\Big) & = \sum_{\Z} \int \int q(\Z, \balpha, \bpi) \ln \{\frac{p(\X, \Z, \balpha, \bpi)}{q(\Z, \balpha, \bpi)}\}d\alpha d\pi \\
    & = \E_{\Z, \balpha, \bpi}[\ln p(\X, \Z, \balpha, \bpi)] - \E_{\Z, \balpha, \bpi}[\ln q(\Z, \balpha, \bpi)] \\
    & = \E_{\Z, \bpi}[\ln p(\X | \Z, \bpi)] + \E_{\Z, \balpha}[\ln p(\Z | \balpha)] + \E_{\balpha}[\ln p(\balpha)] + \E_{\bpi}[\ln p(\bpi)] \\
    &\qquad \:\:\:\:\: -\sum_{i=1}^{N} \E_{\Z_{i}}[\ln q(\Z_{i})] - \E_{\balpha}[\ln q(\balpha)] -  \E_{\bpi}[\ln q(\boldsymbol{\pi)}] \\
    & =  \sum_{i < j}^{N} \sum_{q,l}^{Q}  \tau_{iq} \tau_{jl} \Bigg(X_{ij}
    \Big(\psi(\eta_{ql})  -  \psi(\zeta_{ql})\Big)  + \psi(\zeta_{ql})  -
    \psi(\eta_{ql} + \zeta_{ql})\Bigg) \\
    &\qquad \:\:\:\:\: + \sum_{i=1}^{N} \sum_{q=1}^{Q} \tau_{iq}
    \Big(\psi(n_{q})    -   \psi(\sum_{l=1}^{Q}    n_{l})\Big)    +   \ln
    \Gamma(\sum_{q=1}^{Q}     n_{q}^{0})     -     \sum_{q=1}^{Q}     \ln
    \Gamma(n_{q}^{0}) \\
    &\qquad     \:\:\:\:\:     +     \sum_{q=1}^{Q}\Big(n_{q}^{0} - 1 \Big)\Big(\psi(n_{q})     -
    \psi(\sum_{l=1}^{Q} n_{l}) \Big) + \sum_{q \leq l}^{Q}\Bigg(\ln
    \Gamma(\eta_{ql}^{0} + \zeta_{ql}^{0})\\ 
    &\qquad    \:\:\:\:\:    -    \ln   \Gamma(\eta_{ql}^{0})    -    \ln
    \Gamma(\zeta_{ql}^{0})    +   (\eta_{ql}^{0}    -   1)\Big(\psi(\eta_{ql})
    -\psi(\eta_{ql} + \zeta_{ql})\Big) \\
    &\qquad \:\:\:\:\: +  (\zeta_{ql}^{0}    -   1)\Big(\psi(\zeta_{ql})
    -\psi(\eta_{ql}    +    \zeta_{ql})\Big)\Bigg)    -    \sum_{i=1}^{N}
    \sum_{q=1}^{Q} \tau_{iq} \ln \tau_{iq} \\
    &\qquad   \:\:\:\:\:   -   \ln   \Gamma(\sum_{q=1}^{Q}  n_{q}   )   +
    \sum_{q=1}^{Q} \ln \Gamma(n_{q}) - \sum_{q=1}^{Q}\Big(n_{q} - 1\Big)\Big(\psi(n_{q}) -
    \psi(\sum_{l=1}^{Q} n_{l}) \Big) \\
    &\qquad     \:\:\:\:\:     -     \sum_{q     \leq     l}^{Q}\Bigg(\ln
    \Gamma(\eta_{ql} + \zeta_{ql}) - \ln \Gamma(\eta_{ql}) - \ln
    \Gamma(\zeta_{ql}) \\
    &\qquad   \:\:\:\:\:  +   (\eta_{ql}^{}    -   1)\Big(\psi(\eta_{ql})
    -\psi(\eta_{ql} + \zeta_{ql})\Big)  +  (\zeta_{ql}    -   1)\Big(\psi(\zeta_{ql})
    -\psi(\eta_{ql} + \zeta_{ql})\Big)\Bigg).
  \end{aligned}
  \end{equation}
Therefore
  \begin{equation} \label{equ:lowBound}
    \begin{aligned}
     \mathcal{L}\Big(q(.)\Big) &= \sum_{q < l}^{Q} \Bigg(\Big(\eta_{ql}^{0} - \eta_{ql}
    +       \sum_{i      \neq       j}^{N}       \tau_{iq}      \tau_{jl}
    X_{ij}\Big)\Big(\psi(\eta_{ql})  - \psi(\eta_{ql}  + \zeta_{ql})\Big)
    \\
    &\qquad \:\:\:\:\: + \Big(\zeta_{ql}^{0} - \zeta_{ql}
    +       \sum_{i      \neq       j}^{N}       \tau_{iq}      \tau_{jl}(1-
    X_{ij})\Big)\Big(\psi(\zeta_{ql})       -       \psi(\eta_{ql}      +
    \zeta_{ql})\Big)\Bigg) \\
    &\qquad \:\:\:\:\: + \sum_{q = 1}^{Q} \Bigg(\Big(\eta_{qq}^{0} - \eta_{qq}
    +       \sum_{i     <       j}^{N}       \tau_{iq}      \tau_{jq}
    X_{ij}\Big)\Big(\psi(\eta_{qq})  - \psi(\eta_{qq}  + \zeta_{qq})\Big)
    \\
    &\qquad \:\:\:\:\: + \Big(\zeta_{qq}^{0} - \zeta_{qq}
    +       \sum_{i     <       j}^{N}       \tau_{iq}      \tau_{jq}(1-
    X_{ij})\Big)\Big(\psi(\zeta_{qq})       -       \psi(\eta_{qq}      +
    \zeta_{qq})\Big)\Bigg) \\
    &\qquad   \:\:\:\:\:  +  \sum_{q=1}^{Q}   \Big(n_{q}^{0}  -   n_{q}  +
    \sum_{i=1}^{N} \tau_{iq}\Big)\Big(\psi(n_{q}) -
    \psi(\sum_{l=1}^{Q} n_{l}) \Big) \\
    &\qquad    \:\:\:\:\:    +   \ln\{\frac{\Gamma(\sum_{q=1}^{Q}n_{q}^{0})
      \prod_{q=1}^{Q}\Gamma(n_{q})}{\Gamma(\sum_{q=1}^{Q}n_{q})\prod_{q=1}^{Q}\Gamma(n_{q}^{0})}\}
    +   \sum_{q   \leq   l}^{Q}   \ln   \{\frac{\Gamma(\eta_{ql}^{0}   +
      \zeta_{ql}^{0})\Gamma({\eta_{ql}})\Gamma(\zeta_{ql})}{\Gamma(\eta_{ql})              +
      \zeta_{ql}\Gamma(\eta_{ql}^{0})\Gamma(\zeta_{ql}^{0})}\} \\
    &\qquad \:\:\:\:\: - \sum_{i=1}^{N}\sum_{q=1}^{Q} \tau_{iq} \ln \tau_{iq}.
  \end{aligned}
\end{equation}
After the  variational Bayes  M-step, most of  the terms in  the lower
bound vanish since 
\begin{itemize}
\item $\forall q \: :n_{q} = n_{q}^{0} + \sum_{i=1}^{N} \tau_{iq}$.
 
\item $\forall q \neq l \: :\eta_{ql} = \eta_{ql}^{0} + \sum_{i \neq j}^{N} X_{ij} \tau_{iq} \tau_{jl}$,

\item  $\forall q \: :\eta_{qq} = \eta_{qq}^{0} + \sum_{i < j}^{N} X_{ij} \tau_{iq} \tau_{jq}$.

\item  $\forall q \neq l \: :\zeta_{ql} = \zeta_{ql}^{0} + \sum_{i \neq j}^{N} (1 - X_{ij}) \tau_{iq} \tau_{jl}$,

\item $\forall q \: :\zeta_{qq} = \zeta_{qq}^{0} + \sum_{i < j}^{N} (1
  - X_{ij}) \tau_{iq} \tau_{jq}$.
\end{itemize}
Only the  terms depending  on the   probabilities $\tau_{iq}$
and the normalizing constants of the Dirichlet and Beta distributions remain.

\end{document}